\documentclass[preprintnumbers,article,amsmath,amssymb,floatfix,10pt,prd,onecolumn, superscriptaddress,nofootinbib]{revtex4}
\usepackage[colorlinks=true, pdfstartview=FitV, linkcolor=blue, citecolor=red, urlcolor=magenta]{hyperref}
\usepackage{bbm}
\usepackage{amsfonts}
\usepackage{amsfonts,graphicx,amsmath,amsthm,amssymb,epsfig}
\usepackage{float}
\allowdisplaybreaks
\usepackage{mathrsfs}
\usepackage{latexsym}
\usepackage{epsfig}
\usepackage{epstopdf}
\usepackage{epstopdf}
\usepackage{graphicx}
\usepackage{amssymb}
\usepackage{amsmath}
\usepackage{dcolumn}
\usepackage{bm}
\usepackage{color}
\usepackage{comment}
\usepackage{xcolor}
\begin{document}
\title{\bf Physical validity of anisotropic models derived from isotropic fluid dynamics in
$f(R,T)$ theory: An implication of gravitational decoupling}
\author{Tayyab Naseer}
\email{tayyabnaseer48@yahoo.com}\affiliation{Department of
Mathematics and Statistics, The University of Lahore, 1-KM Defence
Road Lahore-54000, Pakistan} \affiliation{Research Center of
Astrophysics and Cosmology, Khazar University,\\ Baku, AZ1096, 41
Mehseti Street, Azerbaijan}

\author{G. Mustafa}
\email{gmustafa3828@gmail.com}\affiliation{Department of Physics,
Zhejiang Normal University, Jinhua 321004, People's Republic of
China}

\begin{abstract}
In this paper, we derive multiple anisotropic analogs from the
established isotropic model by means of the gravitational decoupling
approach in a fluid-geometry interaction based theory. To accomplish
this, we initially consider a static spherical perfect-fluid
configuration and then introduce a new matter source to induce
anisotropic behavior in the system. The resulting field equations
encapsulate the entire matter distribution and thus become much
complicated. We then split these equations into two sets through
implementing a particular transformation, each set delineating
characteristics attributed to their original fluid sources. We adopt
the Heintzmann's ansatz and some constraints on extra gravitating
source to deal with the first and second systems of equations,
respectively. Furthermore, the two fundamental forms of the matching
criteria are used to make the constant in the considered solution
known. By utilizing the preliminary information of a star candidate
LMC X-4, we assess the physical validity of the developed models.
Our analysis indicates that both our models exhibit characteristics
which are well-agreed with the acceptability criteria for certain
parametric values.
\end{abstract}

\maketitle

\date{\today}

\section{Introduction}

Recent discoveries in cosmology have challenged the traditional view
of the structural configuration of astrophysical bodies existing in
the universe. Instead of randomly distributed in space, they seem to
exhibit systematic organization, which has captured the attention of
researchers. The in-depth examination of interstellar objects has
emerged as a central focus for astronomers aiming to unravel the
enigma behind what we are known with as the accelerated expanding
phenomenon. Dark energy is a mysterious force that appears to be
driving this expansion, counteracting the pull of gravity. The
discovery of dark energy in the late 1990s, through observations of
distant supernovae, was initially met with skepticism but has since
been independently confirmed. Dark energy is estimated to make up
around 68-72$\%$ of the total energy and matter content of the
universe, far exceeding the contributions of normal matter and dark
matter. While general relativity (GR) can explain the expansion of
the universe to some extent, it struggles to account for the
cosmological constant associated with dark energy. This has led
scientists to propose various extensions to GR in an attempt to
better understand the nature of dark energy.

A direct extension within the realm of GR is the $f(R)$ theory,
which stands as a notable progression in theoretical physics. This
theory presents a deviation from the conventional Einstein-Hilbert
action by exchanging the functions of the Ricci scalar $R$ and its
corresponding generic function. Noteworthy for yielding encouraging
outcomes, this theoretical framework has been effectively utilized
in the examination of compact stellar systems \cite{2}-\cite{9}. In
a study by Astashenok and his colleagues \cite{1n}, they explored
the maximum mass limit of stars. Their results suggested that one of
the components in the event GW190814 must be either a fast rotating
neutron star or a massive black hole, effectively eliminating the
possibility of it being a strange star. The existing literature
highlights several significant works \cite{1o,1t}. Bertolami et al.
\cite{10} initially investigated the interaction between fluid and
geometry within the $f(R)$ context. They accomplished this by
combining the matter Lagrangian density and $R$ into a single
function. This concept has prompted scientists to focus their
discussions on the subject of expanding cosmos \cite{11}.

Following this, Harko and co-researchers \cite{20} extended this
idea to the level of action, introducing a novel extension of GR,
known as the $f(R,T)$ theory. In this framework, the interplay
between geometry and the distribution of matter is mediated through
$R$ and trace of the energy-momentum tensor (EMT). This function
gives rise to a non-conservative phenomenon, leading to the
emergence of the additional force (see detail in \cite{22}) within
the field produced by an object, compelling test particles to
deviate from geodesic trajectories. The research conducted by
Houndjo \cite{22a} utilized a minimal gravity model to successfully
explain the transition from a matter-dominated era to a late-time
accelerated cosmic phase. Among the various $f(R,T)$ models, the
$R+2\varpi T$ model has gained significant attention in the
literature as it develops physically feasible interiors. This model
was taken into account by Das et al. \cite{22b} to explore a
three-layers gravastar structure, each layer among them is
corresponded by a different equation of state. Multiple studies have
explored various stellar configurations in the same theory
\cite{c}-\cite{25add}. One key concept within this theory revolves
around its consideration of quantum effects, which in turn open up
possibilities for particle generation. This particular aspect
carries significant importance in the realm of astrophysical
investigations as it suggests a correlation between quantum
principles and $f(R,T)$ gravity. Some important works in this regard
are \cite{1i}-\cite{1l}.

Studying anisotropic models in astrophysics is crucial because many
astrophysical systems, such as neutron stars and other compact
objects, exhibit non-uniform pressure distributions and matter
densities, which cannot be accurately described by isotropic models.
Anisotropy can arise due to various factors, including strong
gravitational fields, rotation, and magnetic fields, leading to
different physical behaviors that are essential for understanding
the stability, structure, and evolution of these celestial bodies.
By incorporating anisotropic models, researchers can gain insights
into phenomena such as the maximum mass of neutron stars and the
effects of pressure anisotropy on surface redshift and thermal
properties. Gravitational decoupling plays a pivotal role in
transforming isotropic fluid models into anisotropic ones by
allowing for the introduction of additional matter sources or fields
that interact differently with the existing fluid dynamics. This
approach facilitates the derivation of new field equations that
reflect the complexities of anisotropic behavior while maintaining a
connection to the original isotropic framework. By employing
transformations and ansatz like Heintzmann's, one can systematically
explore how deviations from isotropy impact the overall dynamics and
stability of stellar configurations. The ability to assess physical
validity through observational data of any star candidate further
enhances the relevance of these models in explaining real
astrophysical phenomena. Thus, the study of anisotropic models not
only enriches our understanding of cosmic structures but also
provides a more accurate representation of the underlying physics
governing these systems.

The standard cosmic model is predominantly founded on the principle
of isotropy and homogeneity on a grand scale. Nevertheless, there
are instances of pressure anisotropy observed at smaller scales
\cite{25ag}-\cite{25ah}. Additionally, the configuration of compact
entities ensures the existence of anisotropy and non-uniform fluid
dispersion. This pressure asymmetry arises from variations in the
radial and transverse directionally-dependent pressure. A group of
elements exists that induces non-uniformity in the internal
structure, including phenomena like phase transitions \cite{25ai},
neutron stars surrounded by intense Maxwell field \cite{25ak}, pion
condensation \cite{25aj}, and various additional influences
\cite{25al,25ala}. An additional contributor to this non-uniformity
is the gravitational impact stemming from tidal forces \cite{25am}.
The uniformity of our cosmos has been investigated through the
analysis of X-ray emissions originating from clusters of galaxies
\cite{25an}. This approach was subsequently applied to large-scale
structures, revealing a conclusion that the universe exhibits
anisotropic properties \cite{25ao}. The presence of pressure
anisotropy is a crucial factor warranting further examination, as it
influences various physical attributes such as gravitational
redshift, mass, etc. within a celestial system.

Astrophysicists have been motivated to seek exact (or numerical)
solutions for celestial structures described by non-linear equations
of motion within GR or alternative theories. The significance of
compact interiors lies in their ability to satisfy requisite
conditions, which necessitates precise solutions. Various
methodologies have been proposed to address this, with gravitational
decoupling emerging as a technique used to discuss solutions
featuring diverse sources such as shear, heat dissipation,
anisotropy, and others. The genesis of this methodology stemmed from
the observation that the field equations encompassing diverse
sources can be disentangled into distinct sets, simplifying the
process of solving each set independently. Ovalle \cite{29}
introduced the minimal geometric deformation (MGD) scheme, a recent
innovation that offers compelling elements for constructing
physically viable stellar solutions, which was extended in
\cite{30}. Additionally, Naseer \cite{31} utilized this approach to
obtain the charged geometry within the complexity-free framework.

Ovalle et al. \cite{33} utilized a spherical isotropic model and
extended it to the anisotropic form by employing the MGD aproach,
ensuring physical feasibility of the solution. In a related
development, Sharif and Sadiq \cite{34} further extended this
technique to include charged configurations, developed two distinct
anisotropic versions of the Krori-Barua solution and examined their
stability. The extensions of multiple geometric solutions has been
made which are based on $f(R)$ formulations \cite{35}. By utilizing
the Durgapal-Fuloria ansatz as a foundational isotropic reference,
multiple physically significant anisotropic analogs have been
obtained \cite{36}. Various researchers have proposed extensions of
isotropic solutions such as Heintzmann and Tolman VII, and
demonstrated their stability within the specified parameter range
\cite{36a,37a}. Researchers considered the axial spacetime and
developed appropriate solutions that exhibit stability \cite{37b}.
Similarly, we have derived charged/uncharged anisotropic analogs of
the Krori-Barua spacetime within a theory possessing strong
non-minimal coupling \cite{37f,37fa}.

We present two distinct models possessing anisotropy by extending
the isotropic spacetime in this article. For this, we incorporate
additional fluid source which may gravitationally coupled to the
primary source within the $f(R,T)$ theory. The subsequent lines
highlights the organizational structure of this paper. It begins
from the next section in which we discuss the foundational aspects
of the modified theory and its corresponding field equations,
focusing on the combined matter source. Some detail on MGD scheme
and how it works is presented in section \textbf{III}. In section
\textbf{IV}, we take into account the Heintzmann's solution and
determine constants involved by this ansatz using junction
conditions. Section \textbf{V} discusses specific conditions that
must be met for the model to be physically acceptable. We present
two novel anisotropic models and their graphical explanation in
section \textbf{VI}. Finally, we provide a summary of our findings
in the concluding section.

\section{Modified $f(R,T)$ Field Equations}

The $f(R,T)$ theory and its implications in the context of
gravitational decoupling can be achieved by modifying the
Einstein-Hilbert action as \cite{20}
\begin{equation}\label{g1}
I=\int \sqrt{-g}\left[\frac{f(R,T)}{16\pi} +L_{m}+\tau
L_{\Phi}\right]d^{4}x,
\end{equation}
in which $L_{m}$ expresses the matter Lagrangian that could be
isotropic, anisotropic, dissipative, etc. according to the scenario
under discussion. Here, we induce the extra gravitating source with
the seed source whose Lagrangian is indicated by $L_{\Phi}$. The
quantity $\tau$ is referred to the decoupling parameter, commanding
the impact of extra source on the original gravitational field. The
action \eqref{g1}, after some lengthy calculations, produce the
following governing equations as
\begin{equation}\label{g2}
G_{\sigma\omega}=8\pi T_{\sigma\omega}^{\textsf{(tot)}},
\end{equation}
where the factors on left and right sides explain the geometrical
structure and the fluid distribution in the celestial interior,
respectively. We categorize the term
$T_{\sigma\omega}^{\textsf{(tot)}}$ as
\begin{equation}\label{g3}
T_{\sigma\omega}^{\textsf{(tot)}}=T_{\sigma\omega}^{\textsf{(eff)}}+\tau
\Phi_{\sigma\omega}=\frac{1}{f_{R}}T_{\sigma\omega}+T_{\sigma\omega}^{\textsf{(cor)}}+\tau
\Phi_{\sigma\omega}.
\end{equation}
Here,
\begin{itemize}
\item $\Phi_{\sigma\omega}$ indicates the extra matter source,
\item $T_{\sigma\omega}$ and $T_{\sigma\omega}^{\textsf{(cor)}}$ are
usual EMT and corrections due to modification of gravity. These
corrections are defined by
\begin{eqnarray}
\nonumber T_{\sigma\omega}^{\textsf{(cor)}}&=&\frac{1}{8\pi
f_{R}}\bigg[f_{T}T_{\sigma\omega}+\bigg\{\frac{R}{2}\bigg(\frac{f}{R}
-f_{R}\bigg)-L_{m}f_{T}\bigg\}g_{\sigma\omega}\\\label{g4}
&-&(g_{\sigma\omega}\Box-\nabla_{\sigma}\nabla_{\omega})f_{R}+2f_{T}g^{\zeta\beta}\frac{\partial^2
L_{m}}{\partial g^{\sigma\omega}\partial g^{\zeta\beta}}\bigg],
\end{eqnarray}
\end{itemize}
where $f_{T}=\frac{\partial f}{\partial T}$ and
$f_{R}=\frac{\partial f}{\partial R}$. Furthermore, the D'Alembert
operator $\Box$ is defined by
$\Box\equiv\frac{1}{\sqrt{-g}}\partial_\sigma\big(\sqrt{-g}g^{\sigma\omega}\partial_{\omega}\big)$
and we symbolize the covariant derivative by $\nabla_\sigma$.
Following recent literature regarding the compact stellar solutions,
we choose $L_{m}=P$ in this case with $P$ being the isotropic
pressure.

The isotropic EMT describes a system where the energy and momentum
densities are the same in all spatial directions. This is a crucial
assumption in modeling the large-scale structure of the universe,
which is observed to be approximately homogeneous and isotropic on
the largest scales. This is given as follows
\begin{equation}\label{g5}
T_{\sigma\omega}=(\mu+P)\textsf{U}_{\sigma}\textsf{U}_{\omega}+P
g_{\sigma\omega},
\end{equation}
where
\begin{itemize}
\item $\mu$ being the density of the fluid,
\item $\textsf{U}_{\sigma}$ is the four-velocity.
\end{itemize}
The field equations of this modified gravity theory have the
following trace component as
\begin{align}\nonumber
&2f+T(f_T+1)-Rf_R-3\nabla^{\sigma}\nabla_{\sigma}f_R-4f_TL_m
+2f_Tg^{\zeta\beta}g^{\sigma\omega}\frac{\partial^2L_m}{\partial
g^{\zeta\beta}\partial g^{\sigma\omega}}=0.
\end{align}
It must be highlighted the connection between different theories of
gravity under certain circumstances. For instance, vanishing
coupling of matter with geometry leads these results to $f(R)$
framework. The non-null divergence of the $f(R,T)$ EMT indicates a
departure from the conservation phenomenon, leading to the emergence
of an additional force affecting the trajectories of test particles
moving in their gravitational fields. We can write it mathematically
as
\begin{align}\nonumber
\nabla^\sigma
T_{\sigma\omega}&=\frac{f_T}{8\pi-f_T}\bigg[(g_{\sigma\omega}L_m-T_{\sigma\omega})\nabla^\sigma\ln{f_T}
-\frac{1}{2}g_{\zeta\beta}\nabla_\omega T^{\zeta\beta}\\\label{g11}
&-\frac{8\pi\tau}{f_T}\nabla^\sigma\Phi_{\sigma\omega}
+\nabla^\sigma(g_{\sigma\omega}\pounds_m-2\mathbb{T}_{\sigma\omega})\bigg].
\end{align}

The spherical interior metric plays a crucial role in the study of
compact objects like black holes and neutron stars, providing a
mathematical framework to describe the internal structure of these
astrophysical entities. This spacetime is represented by the line
element given by
\begin{equation}\label{g6}
ds^2=-e^{p_1} dt^2+e^{p_2} dr^2+r^2\big(d\theta^2+\sin^2\theta
d\phi^2\big),
\end{equation}
where $p_1=p_1(r)$ and $p_2=p_2(r)$. The four-velocity defined
earlier in Eq.\eqref{g5} is now become
\begin{equation}\label{g7}
\textsf{U}_\sigma=-\delta^0_\sigma
e^{\frac{p_1}{2}}=(-e^{\frac{p_1}{2}},0,0,0).
\end{equation}
Adopting specific models allows for more flexibility in describing
the gravitational interactions and the cosmic evolution. Also, by
doing so, we can construct solutions that are better able to match
the observational data. Following is a minimal $f(R,T)$ model with a
real-valued constant $\varrho$ as
\begin{equation}\label{g7a}
f(R,T)=f_1(R)+ f_2(T)=R+2\varrho T.
\end{equation}
The reason behind choosing this linear model is that we shall apply
the decoupling strategy in the next section, requiring the field
equations to be in simplest form so that they can easily be handled.
Using the inflation potentials in their work, Ashmita et al.
\cite{38} obtained some significant results and attempted to match
them with the existing data. They came to the conclusion that their
results are consistent only for a particular range of the model
constant, i.e., $-0.37<\varrho<1.483$. Some other considerable works
in this context are \cite{39}-\cite{41}.

Using Eqs.\eqref{g2}, \eqref{g6} and \eqref{g7a} together leads to
the development of the field equations as
\begin{align}\label{g8}
&e^{-p_2}\left(\frac{p_2'}{r}-\frac{1}{r^2}\right)
+\frac{1}{r^2}=8\pi\left(\mu-\tau\Phi_{0}^{0}\right)+\varrho\left(3\mu-P\right),\\\label{g9}
&e^{-p_2}\left(\frac{1}{r^2}+\frac{p_1'}{r}\right)
-\frac{1}{r^2}=8\pi\left(P+\tau\Phi_{1}^{1}\right)-\varrho\left(\mu-3P\right),
\\\label{g10}
&\frac{e^{-p_2}}{4}\left[p_1'^2-p_2'p_1'+2p_1''-\frac{2p_2'}{r}+\frac{2p_1'}{r}\right]
=8\pi\left(P+\tau\Phi_{2}^{2}\right)-\varrho\left(\mu-3P\right),
\end{align}
along with the non-conservation equation \eqref{g11} expressed by
\begin{align}\nonumber
&\frac{dP}{dr}+\frac{p_1'}{2}\left(\mu+P\right)+\frac{\tau p_1'}{2}
\left(\Phi_{1}^{1}-\Phi_{0}^{0}\right)+\tau\frac{d\Phi_{1}^{1}}{dr}\\\label{g12}
&+\frac{2\tau}{r}\left(\Phi_{1}^{1}-\Phi_{2}^{2}\right)=-\frac{\varrho}{4\pi-\varrho}\big(\mu'-P'\big).
\end{align}
Equation \eqref{g12} describes that the sum of all internal/external
forces acting on a celestial body must be null to maintain the
equilibrium, thus it is necessarily needed to explore this
phenomenon in the presence of correction terms. The solution to a
set \eqref{g8}-\eqref{g10} become obscured due to high degrees of
freedom, such as
$(p_1,p_2,\mu,P,\Phi_{0}^{0},\Phi_{1}^{1},\Phi_{2}^{2})$. In such
cases, it becomes necessary to choose certain constraints to ensure
a unique solution.

\section{Exploring Gravitational Decoupling via MGD}

Since we discuss the need of a systematic scheme in the previous
section, the gravitational decoupling strategy serves as a best
candidate to deal with such problems \cite{33}. This scheme helps to
simplify the complicated field equations by transforming them into
another reference frame. Its implementation can be done only if a
new metric is considered as a perfect-fluid solution to
Eqs.\eqref{g8}-\eqref{g10}. We adopt it as
\begin{equation}\label{g15}
ds^2=-e^{p_3(r)}dt^2+\frac{1}{p_4(r)}dr^2+r^2\big(d\theta^2+\sin^2\theta
d\phi^2\big).
\end{equation}
The equations help to transform both metric components are expressed
as
\begin{equation}\label{g16}
p_3\rightarrow p_1=p_3+\tau\mathrm{d}_1, \quad p_4\rightarrow
e^{-p_2}=p_4+\tau\mathrm{d}_2.
\end{equation}
In two types of the decoupling technique \cite{aa,aaa}, we adopt the
MGD scheme which ensures the transformation of only the radial
metric component. Following this, the new form of Eq.\eqref{g16} is
\begin{equation}\label{g17}
p_3\rightarrow p_1=p_3, \quad p_4\rightarrow
e^{-p_2}=p_4+\tau\mathrm{d},
\end{equation}
where $\mathrm{d}=\mathrm{d}(r)$. The beneficial point of these
transformations is that they always preserve the spherical symmetry.
We choose the MGD approach for the current study due to its
effectiveness in addressing the complexities associated with
gravitational decoupling in anisotropic models. There are many
advantages of MGD discussed below.
\begin{itemize}
\item The MGD approach simplifies the
process of obtaining exact solutions by focusing on minimal
modifications to the geometric structure of spacetime. This makes it
easier to derive anisotropic analogs from isotropic models without
introducing unnecessary complications.
\item The MGD provides a clear framework for
interpreting the physical implications of the derived models,
ensuring that the solutions remain consistent with known
astrophysical phenomena.
\item Since our study is rooted in
fluid-geometry interactions, MGD aligns well with the underlying
principles of fluid dynamics, facilitating a seamless transition
from isotropic to anisotropic configurations.
\item Alternative methods for
gravitational decoupling often involve more complex transformations
or assumptions that may not be necessary for our specific context.
These methods can sometimes lead to solutions that are less
physically intuitive or harder to interpret in terms of
astrophysical applications.
\end{itemize}

Implementing the transformation \eqref{g17} on
Eqs.\eqref{g8}-\eqref{g10} split them form different choices of
$\eta$. The first set where the effect of additional source is
disappeared, i.e., $\tau=0$, is obtained as
\begin{align}\label{g18}
&e^{-p_2}\left(\frac{p_2'}{r}-\frac{1}{r^2}\right)
+\frac{1}{r^2}=8\pi\mu+\varrho\left(3\mu-P\right),\\\label{g19}
&e^{-p_2}\left(\frac{1}{r^2}+\frac{p_1'}{r}\right)
-\frac{1}{r^2}=8\pi P-\varrho\left(\mu-3P\right),\\\label{g20}
&\frac{e^{-p_2}}{4}\left[p_1'^2-p_2'p_1'+2p_1''-\frac{2p_2'}{r}+\frac{2p_1'}{r}\right]
=8\pi P-\varrho\left(\mu-3P\right).
\end{align}
It is enough to adopt first two equations from the above set to
determine the fluid parameters in their explicit form defined by
\begin{align}\label{g18a}
\mu&=\frac{e^{-p_2}}{8r^2\big(\varrho^2+6\pi\varrho+8\pi^2\big)}\big[\varrho
r p_1 '+(3 \varrho +8 \pi ) r p_2 '+2 (\varrho +4 \pi ) \big(e^{p_2
}-1\big)\big],\\\label{g19a}
P&=\frac{e^{-p_2}}{8r^2\big(\varrho^2+6\pi\varrho+8\pi^2\big)}\big[(3
\varrho +8 \pi ) r p_1 '+\varrho  r p_2 '-2 (\varrho +4 \pi )
\big(e^{p_2 }-1\big)\big].
\end{align}

As the second set completely characterizing the extra fluid source
is concerned, we find the following field equations
\begin{align}\label{g21}
&8\pi\Phi_{0}^{0}=\frac{\mathrm{d}'}{r}+\frac{\mathrm{d}}{r^2},\\\label{g22}
&8\pi\Phi_{1}^{1}=\mathrm{d}\left(\frac{p_1'}{r}+\frac{1}{r^2}\right),\\\label{g23}
&8\pi\Phi_{2}^{2}=\frac{\mathrm{d}}{4}\left(2p_1''+p_1'^2+\frac{2p_1'}{r}\right)
+\mathrm{d}'\left(\frac{p_1'}{4}+\frac{1}{2r}\right).
\end{align}
Another fact about MGD scheme is the conservation of both fluid
setups independently, leading to the phenomenon of not exchanging
the energy between them. Since there are four terms
($\mu,P,p_1,p_2$) in Eqs.\eqref{g18a} and \eqref{g19a} which need to
be calculated, we may consider an ansatz to deal with these
equations. Further, the same number of unknowns
($\mathrm{d},\Phi_{0}^{0},\Phi_{1}^{1},\Phi_{2}^{2}$) are present in
Eqs.\eqref{g21}-\eqref{g23}, a single constraint on $\Phi$-sector is
needed at a time. After adding the additional source, the new
variables must be defined. This is done in the following
\begin{equation}\label{g13}
\hat{\mu}=\mu-\tau\Phi_{0}^{0},\quad \hat{P}_{r}=P+\tau\Phi_{1}^{1},
\quad \hat{P}_{t}=P+\tau\Phi_{2}^{2},
\end{equation}
along with the total anisotropy given by
\begin{equation}\label{g14}
\hat{\Delta}=\hat{P}_{t}-\hat{P}_{r}=\tau(\Phi_{2}^{2}-\Phi_{1}^{1}),
\end{equation}
vanishes when $\tau=0$. In other words, when the effects of the
extra source are extracted, we are left with the initial perfect
fluid distribution.

\section{Heintzmann's Spacetime and Junction Conditions}

We devote this section for the development of a unique solution to
Eqs.\eqref{g18a} and \eqref{g19a}. The need for a particular metric
component based solution is already discussed earlier. Following
this, a recent solution has been proposed in the literature that
helps in exploring the physics of physically relevant stellar
structures \cite{42a1,42a2}. This perfect solution has the form
\begin{align}\label{g33}
e^{p_1(r)}&=b_1^2 \left(b_2 r^2+1\right)^3,\\\label{g34}
e^{p_2(r)}&=\frac{2 \left(b_2 r^2+1\right)\sqrt{4 b_2
r^2+1}}{\left(2-b_2 r^2\right)\sqrt{4 b_2 r^2+1}-3 b_2 b_3
r^2},\\\nonumber \mu&=\frac{3b_2}{4 (\alpha +2 \pi ) (\alpha +4 \pi
) \big(b_2 r^2+1\big)^2 \big(4 b_2 r^2+1\big)^{3/2}}\big[2 (\alpha
+3 \pi ) b_3+2 b_2 r^2\\\nonumber &\times \big\{(2 \alpha +9 \pi )
b_3+(6 \alpha +13 \pi ) \sqrt{4 b_2 r^2+1}\big\}+3 (\alpha +2 \pi )
\sqrt{4 b_2 r^2+1}\\\label{g35} &+b_2^2 r^4 \big(8 \pi  \sqrt{4 b_2
r^2+1}-7 \alpha b_3\big)\big],\\\nonumber P&=\frac{-3b_2}{4 (\alpha
+2 \pi ) (\alpha +4 \pi ) \big(b_2 r^2+1\big)^2 \big(4 b_2
r^2+1\big)^{3/2}}\bigg[2 b_2 r^2 \big\{(3 \alpha +11 \pi )
b_3\\\nonumber &-(5 \alpha +9 \pi ) \sqrt{4 b_2 r^2+1}\big\}-3
(\alpha +2 \pi ) \sqrt{4 b_2 r^2+1}+b_2^2 r^4 \big\{7 (3 \alpha +8
\pi ) b_3\\\label{g36} &+8 (\alpha +3 \pi ) \sqrt{4 b_2
r^2+1}\big\}+2 \pi b_3\bigg].
\end{align}

The advantage to utilize Heintzmann's ansatz in our analysis is
based on several key factors that enhance the robustness and
applicability of our results. They are described as
\begin{itemize}
\item This ansatz is well-established in
the literature for modeling anisotropic stellar structures. Its
previous successful applications provide a strong foundation for our
work and lend credibility to our chosen methodology.
\item This ansatz allows for a
straightforward mathematical treatment of the field equations,
making it easier to derive explicit solutions while ensuring that
the resulting models retain essential physical characteristics.
\item While Heintzmann's ansatz is
specific, it does not significantly limit the generality of our
findings. The solutions derived from this approach can be extended
or modified by varying parameters or considering different forms of
matter sources. This adaptability ensures that our results remain
relevant across a range of astrophysical scenarios.
\end{itemize}

Equations \eqref{g33} and \eqref{g34} contain a triplet of constants
$(b_1,b_2,b_3)$ that needs to be known so that the graphical
analysis can be performed later for our resulting solutions. To
perform this analysis, some conditions on the spherical interface
($\Sigma:~r=\bar{R}$) have been proposed, referred to the junction
conditions. We need both interior and exterior metrics to use such
conditions. The former metric has the form for components
\eqref{g33} and \eqref{g34} as
\begin{align}\nonumber
ds_-^2&=b_1^2 \left(b_2 r^2+1\right)^3dt^2\\\label{g36aa} &+\frac{2
\left(b_2 r^2+1\right)\sqrt{4 b_2 r^2+1}}{\left(2-b_2
r^2\right)\sqrt{4 b_2 r^2+1}-3 b_2 b_3 r^2}
dr^2+r^2\left(d\theta^2+\sin^2\theta d\phi^2\right),
\end{align}
while the later metric is adopted as the Schwarzschild solution with
total mass symbolized by $\bar{M}$ given by
\begin{equation}\label{g25}
ds_+^2=-\frac{r-2\bar{M}}{r}dt^2+\frac{r}{r-2\bar{M}}dr^2+
r^2\left(d\theta^2+\sin^2\theta d\phi^2\right).
\end{equation}
According to continuity of the first fundamental form, the following
expressions hold at the boundary as
\begin{equation}\nonumber
g_{tt}^-~{_=^\Sigma}~g_{tt}^+, \quad g_{rr}^-~{_=^\Sigma}~g_{rr}^+,
\end{equation}
resulting in when combined with Eqs.\eqref{g36aa} and \eqref{g25} as
\begin{eqnarray}\label{g36ab}
\frac{\bar{R}-2\bar{M}}{\bar{R}}&=&b_1^2\left(b_2\bar{R}^2+1\right)^3,\\\label{g36ac}
\frac{\bar{R}}{\bar{R}-2\bar{M}}&=&\frac{2 \left(b_2
\bar{R}^2+1\right)\sqrt{4 b_2 \bar{R}^2+1}}{\left(2-b_2
\bar{R}^2\right)\sqrt{4 b_2 \bar{R}^2+1}-3 b_2 b_3 \bar{R}^2}.
\end{eqnarray}

Another constraint achieved from the second fundamental form is the
disappearance of the pressure at the hypersurface, i.e.,
$P~{_=^\Sigma}~0$. This condition yields when used in Eq.\eqref{g36}
as
\begin{align}\nonumber
&2 b_2 \bar{R}^2 \big\{(3 \alpha +11 \pi ) b_3-(5 \alpha +9 \pi )
\sqrt{4 b_2 \bar{R}^2+1}\big\}-3 (\alpha +2 \pi ) \sqrt{4 b_2
\bar{R}^2+1}\\\label{g36ad} &+b_2^2 \bar{R}^4 \big\{7 (3 \alpha +8
\pi ) b_3+8 (\alpha +3 \pi ) \sqrt{4 b_2 \bar{R}^2+1}\big\}+2 \pi
b_3 =0.
\end{align}
We attempt to solve Eqs.\eqref{g36ab}-\eqref{g36ad} simultaneously
to get three constants but find it not possible. Therefore, we
consider $b_2$ as a free parameter and calculate the remaining
quantities in terms of this factor by using Eqs.\eqref{g36ab} and
\eqref{g36ad} as
\begin{align}\label{g37}
b_{1}&=\sqrt{\frac{\bar{R}-2 \bar{M}}{b_2^3 \bar{R}^7+3 b_2^2
\bar{R}^5+3 b_2 \bar{R}^3+\bar{R}}},\\\label{g38a}
b_{3}&=\frac{\sqrt{4 b_2 \bar{R}^2+1} \left(3 \varrho -8 \varrho
b_2^2 \bar{R}^4-24 \pi  b_2^2 \bar{R}^4+10 \varrho  b_2 \bar{R}^2+18
\pi b_2 \bar{R}^2+6 \pi \right)}{21 \varrho  b_2^2 \bar{R}^4+56 \pi
b_2^2 \bar{R}^4+6 \varrho b_2 \bar{R}^2+22 \pi  b_2 \bar{R}^2+2\pi}.
\end{align}
\begin{table}[H]
\scriptsize \centering \caption{Constant doublet ($b_{1},b_{3}$) for
$b_{2}=0.001~km^{-2}$ corresponding to multiple choices of
$\varrho$.} \label{Table1} \vspace{+0.07in}
\setlength{\tabcolsep}{0.95em}
\begin{tabular}{ccccccccccc}
\hline\hline $\varrho$ & 0 & 0.2 & 0.4 & 0.6 & 0.8 & 1.0 & 1.2
\\\hline $b_{1}$ & 0.718457 & 0.718457 & 0.718457 & 0.718457 &
0.718457 & 0.718457 & 0.718457
\\\hline $b_{3}$ & 2.1283 & 2.1789 & 2.2286 & 2.2775 &
2.3256 & 2.3729 & 2.4194\\
\hline\hline
\end{tabular}
\end{table}
\begin{table}[H]
\scriptsize \centering \caption{Constant doublet ($b_{1},b_{3}$) for
$b_{2}=0.003~km^{-2}$ corresponding to multiple choices of
$\varrho$.} \label{Table2} \vspace{+0.07in}
\setlength{\tabcolsep}{0.95em}
\begin{tabular}{ccccccccccc}
\hline\hline $\varrho$ & 0 & 0.2 & 0.4 & 0.6 & 0.8 & 1.0 & 1.2
\\\hline $b_{1}$ & 0.598964 & 0.598964 & 0.598964 & 0.598964 &
0.598964 & 0.598964 & 0.598964
\\\hline $b_{3}$ & 1.3145 & 1.3396 & 1.3639 & 1.3876 &
1.4105 & 1.4328 & 1.4545\\
\hline\hline
\end{tabular}
\end{table}
\begin{table}[H]
\scriptsize \centering \caption{Constant doublet ($b_{1},b_{3}$) for
$b_{2}=0.005~km^{-2}$ corresponding to multiple choices of
$\varrho$.} \label{Table3} \vspace{+0.07in}
\setlength{\tabcolsep}{0.95em}
\begin{tabular}{ccccccccccc}
\hline\hline $\varrho$ & 0 & 0.2 & 0.4 & 0.6 & 0.8 & 1.0 & 1.2
\\\hline $b_{1}$ & 0.509276 & 0.509276 & 0.509276 & 509276 &
0.509276 & 0.509276 & 0.509276
\\\hline $b_{3}$ & 0.8888 & 0.9059 & 0.9223 & 0.9382 &
0.9536 & 0.9684 & 0.9828\\
\hline\hline
\end{tabular}
\end{table}
\noindent For different values of $b_2$ and the model parameter, we
find the two constants $b_1$ and $b_3$ in Tables
\textbf{I}-\textbf{III}. Note that these numerical values are
obtained by using the approximate mass and radius of a compact star
LMC X-4 as $\bar{M}=1.04 \pm 0.09 M_{\bigodot}$ and $\bar{R}=8.301
\pm 0.2 km$ \cite{42aa}. This star is a well-studied binary system
that contains a neutron star, making it an ideal candidate for
examining the physical properties of compact stars. Its
characteristics, including its mass, radius, and behavior under
various conditions, have been extensively documented in the
literature. This wealth of observational data allows us to make
informed comparisons between our theoretical models and the actual
physical properties of LMC X-4. Moreover, this star exhibits
significant variability in its X-ray emissions and has been
associated with various astrophysical phenomena, such as pulsations
and outbursts. These features provide a rich context for testing the
implications of our derived models. We observe from these Tables
that the term $b_1$ is independent of $\varrho$, however, decreases
with the rise in $b_2$. On the other hand, the quantity $b_3$ in
directly and inversely related with $\varrho$ and $b_2$,
respectively.

\section{Physical Requirements for a Viable Stellar Model}

The solution of the Einstein or modified equations of motion shall
represent a physically existing compact interior only if it is
consistent with multiple conditions regarding acceptability of that
model. Such conditions have been suggested and interrogated in the
scientific literature \cite{ab}-\cite{ad}. Some highlights can be
found in the following.
\begin{itemize}
\item Preventing from the existence of singularities in geometric
terms at any point in the interior is one of the basic pillar of the
solution to be physically feasible. Hence, the metric components, we
choose or find, must obey increasing profile w.r.t. $r$ and no
singular point should occur anywhere. To check this, we achieve the
coefficients \eqref{g33} and \eqref{g34} for $r=0$ as
\begin{align}\nonumber
e^{p_1(r)}|_{r=0}=b_1^2, \quad e^{p_2(r)}|_{r=0}=1,
\end{align}
indicating non-singularity. Also, the first-order derivatives of the
same components are
\begin{align}\nonumber
(e^{p_1(r)})'&=6 b_1^2 b_2 r \left(b_2 r^2+1\right)^2,\\\nonumber
(e^{p_2(r)})'&=\frac{12 b_2 r \left\{2 b_2 r^2 \left(b_3 \left(1-b_2
r^2\right)+2 \sqrt{4 b_2 r^2+1}\right)+\sqrt{4 b_2
r^2+1}+b_3\right\}}{\sqrt{4 b_2 r^2+1} \left\{b_2 r^2 \left(\sqrt{4
b_2 r^2+1}+3 b_3\right)-2 \sqrt{4 b_2 r^2+1}\right\}^2}.
\end{align}
It is obvious from above equations that when we substitute $r=0$,
both derivatives become null and negative, otherwise. This confirms
that the metric potentials possess an increasing behavior. We also
check this graphically, but not add their plots here.

\item The energy density should be typically positive and finite throughout the stellar region.
The radial/tangential pressures also exhibit positive and finite
values within the compact star, with the tangential pressure often
being different from the radial one, indicating the presence of
anisotropy. Further, they all must be maximum and minimum at the
stellar core and boundary, respectively.

\item The spherical mass function is a crucial element to be discussed while studying compact stars.
This is because this factor allows for the investigation of various
physical characteristics, such as the surface redshift,
gravitational lensing, etc. Its mathematical formula is \cite{42a3}
\begin{equation}\label{g39}
\hat{m}(r)=\frac{1}{2}\int_{0}^{\bar{R}}\bar{r}^2\mu d\bar{r}.
\end{equation}
It is interesting to know the closeness of particles within a
compact star to explore how dense it is. In other words, the
compactness is equal to the mass-radius ratio, symbolically
expresses as
\begin{equation}\label{g40}
\lambda(r)=\frac{\hat{m}(r)}{r}.
\end{equation}
A research done by Buchdahl shows its upper limit in the case of
spherical fluid distribution as $\frac{4}{9}$ \cite{42a}. Another
factor, named the redshift (or surface redshift) can be defined in
relation with the mass as
\begin{equation}\label{g41}
Z_{rs}(r)=\frac{1-\sqrt{1-2\lambda(r)}}{\sqrt{1-2\lambda(r)}}.
\end{equation}
Different studies have been done in order to explore its acceptable
values inside a compact interior and it was observed to be not
greater than $5.211$ \cite{42b}.

\item The energy conditions play a crucial role in the study of compact
stars as they provide essential constraints on the physical
properties of these astrophysical objects. By validating such
bounds, we can ensure the consistency of theoretical models with the
fundamental principles of relativity and the behavior of matter
under extreme gravitational conditions. They are, in this case,
given by \cite{42bc,42bd}
\begin{equation}
\left.
\begin{aligned}\label{g50}
&\hat{\mu} \geq 0, \quad \hat{\mu}+\hat{P}_{t} \geq 0, \\
&\hat{\mu}+\hat{P}_{r} \geq 0, \quad \hat{\mu}-\hat{P}_{t} \geq 0,\\
&\hat{\mu}-\hat{P}_{r} \geq 0, \quad
\hat{\mu}+2\hat{P}_{t}+\hat{P}_{r} \geq 0.
\end{aligned}
\right\}
\end{equation}
\item As the stability of massive systems is concerned, the models which fulfil stability criteria
are more likely to be studied for further exploration. In this
regard, the literature suggests multiple schemes that can be used to
analyze whether a resulting model is stable or not. For example, the
causality condition is a well-established way to do so \cite{42bb}.
According to this, the light's speed must be greater than that of
the sound, i.e., $$0 <
\hat{V}_{t}^{2}=\frac{d\hat{P}_{t}}{d\hat{\mu}} < 1, \quad 0 <
\hat{V}_{r}^{2}=\frac{d\hat{P}_{r}}{d\hat{\mu}} < 1.$$

Further, cracking can occur in a compact star due to sensitivity of
the radial forces to local density perturbations. This can cause the
self-gravitating object to exhibit instability within certain ranges
of some physical parameters. Hence, Herrera \cite{42ba} suggested a
condition that must be fulfilled to avoid the cracking as $0 <
|\hat{V}_{t}^{2}-\hat{V}_{r}^{2}| < 1$.
\end{itemize}

\section{Modeling Anisotropic Solutions}

This section deals with the solution of second set
\eqref{g21}-\eqref{g23}. For this, we use multiple constraints to
make highly complicated differential equations easy to solve. The
following subsections shall present different models and their
physical acceptability.

\subsection{Model 1: Density-like Constraint}

To obtain the first model, we consider that the density of perfect
fluid is equal to that of the additional source. This constraint has
widely been employed in the literature while studying compact
systems. In terms of mathematical symbols, we can write it as
\cite{42baba}
\begin{equation}\label{g51}
\mu=\Phi_{0}^{0}.
\end{equation}
Equation \eqref{g51} leads to a linear-order differential equation
when we combine it with \eqref{g18a} and \eqref{g22} as
\begin{align}\nonumber
&\frac{1}{8\pi}\bigg\{\frac{\mathrm{d}'(r)}{r}+\frac{\mathrm{d}(r)}{r^2}\bigg\}
-\frac{e^{-p_2}}{8r^2\big(\varrho^2+6\pi\varrho+8\pi^2\big)}\\\label{g52}
&\times\big[\varrho r p_1 '+(3 \varrho +8 \pi ) r p_2 '+2 (\varrho
+4 \pi ) \big(e^{p_2 }-1\big)\big]=0.
\end{align}
Writing the above equation in terms of Heintzmann's potentials
\eqref{g33} and \eqref{g34}, we are left with
\begin{align}\nonumber
&\frac{1}{8\pi}\bigg\{\frac{\mathrm{d}'(r)}{r}+\frac{\mathrm{d}(r)}{r^2}\bigg\}-\frac{3b_2}{4
(\alpha +2 \pi ) (\alpha +4 \pi ) \psi ^3 \big(b_2
r^2+1\big)^2}\bigg[3 (\varrho +2 \pi ) \psi +2b_3
\\\label{g53} &\times(\varrho +3 \pi )+b_2^2 r^4 \big(8 \pi  \psi -7 \varrho
b_3\big)+2 b_2 r^2 \big\{(6 \varrho +13 \pi ) \psi +(2 \varrho +9
\pi ) b_3\big\}\bigg]=0,
\end{align}
where $\psi=\sqrt{4 b_2 r^2+1}$. We can find the term
$\mathrm{d}(r)$ from the above equation, either numerically or
analytically. So we first attempt to calculate the exact solution
and become able to achieve this. The obtained form for
$\mathrm{d}(r)$ is given by
\begin{align}\nonumber
\mathrm{d}(r)&=\frac{B_1}{r}+\frac{3 \pi }{4 \big(\varrho ^2+6 \pi
\varrho +8 \pi ^2\big) \sqrt{b_2} r}\bigg[\frac{2 \sqrt{b_2} b_3 r
\big(\varrho +(7 \varrho +8 \pi ) b_2 r^2\big)}{\psi  \big(b_2
r^2+1\big)}\\\nonumber &-\frac{4 (3 \varrho +4 \pi ) \sqrt{b_2}
r}{b_2 r^2+1}+4 \sqrt{3} \varrho  b_3 \tanh
^{-1}\bigg(\frac{\sqrt{3} \sqrt{b_2} r}{\psi }\bigg)+16 \pi
\sqrt{b_2} r\\\label{g53} &-7 \varrho  b_3 \sinh ^{-1}\big(2
\sqrt{b_2} r\big)+12 \varrho \tan ^{-1}\big(\sqrt{b_2} r\big)\bigg],
\end{align}
with an integration constant $B_1$ whose value must be taken as zero
to avoid the singularity at $r=0$. Further, the corresponding
$g_{rr}$ potential is determined using Eq.\eqref{g17} as
\begin{align}\nonumber
e^{p_2(r)}&=\frac{2 \left(b_2 r^2+1\right)\sqrt{4 b_2
r^2+1}}{\left(2-b_2 r^2\right)\sqrt{4 b_2 r^2+1}-3 b_2 b_3
r^2}+\frac{3 \pi \tau }{4 \big(\varrho ^2+6 \pi \varrho +8 \pi
^2\big) \sqrt{b_2} r}\\\nonumber &\times\bigg[\frac{2 \sqrt{b_2} b_3
r \big(\varrho +(7 \varrho +8 \pi ) b_2 r^2\big)}{\psi  \big(b_2
r^2+1\big)}-\frac{4 (3 \varrho +4 \pi ) \sqrt{b_2} r}{b_2 r^2+1}+16
\pi \sqrt{b_2} r+4 \sqrt{3}\\\label{g54} &\times \varrho  b_3 \tanh
^{-1}\bigg(\frac{\sqrt{3b_2} r}{\psi }\bigg)-7 \varrho  b_3 \sinh
^{-1}\big(2 \sqrt{b_2} r\big)+12 \varrho \tan ^{-1}\big(\sqrt{b_2}
r\big)\bigg].
\end{align}
We can use this along with Eqs.\eqref{g13} and \eqref{g14} to find
the final expressions of effective matter determinants. Certain
characteristics corresponding to this developed model are now ready
to be investigated in this modified gravity context. For this, we
need to choose some parametric values so that the graphical analysis
can be smoothly performed. In this regard, we adopt
$\tau=0.1,0.2,0.3,~ \varrho=0.1,0.6,1.1$ and $b_2=0.005~km^{-2}$ to
observe how the considered minimal model \eqref{g7a} and decoupling
strategy affect our results. Figure \textbf{1} shows the graphs of
the deformation function \eqref{g53} and modified potential
\eqref{g54} for all above parametric choices. Both plots show
consistent data within their domains.
\begin{figure}[h!]\center
\epsfig{file=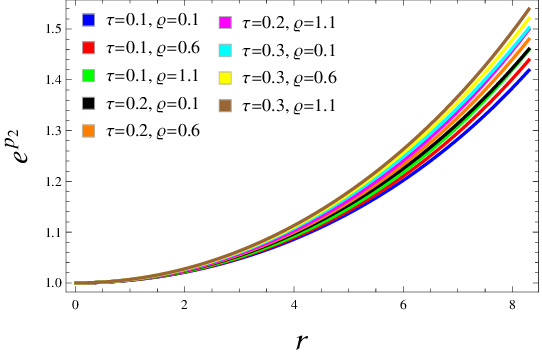,width=0.4\linewidth}\epsfig{file=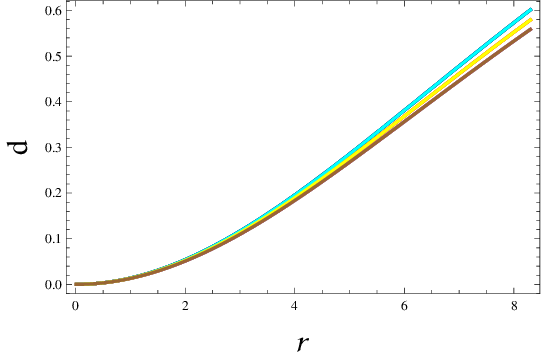,width=0.4\linewidth}
\caption{Modified $g_{rr}$ potential and deformation function for
model 1.}
\end{figure}
\begin{figure}[h!]\center
\epsfig{file=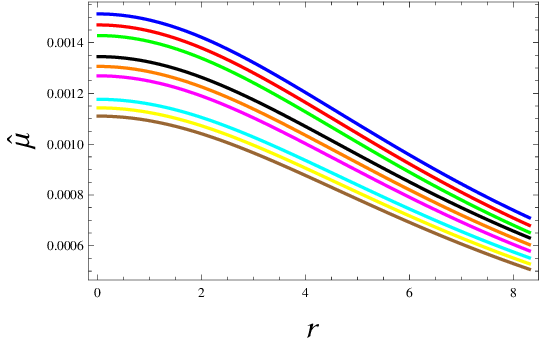,width=0.4\linewidth}\epsfig{file=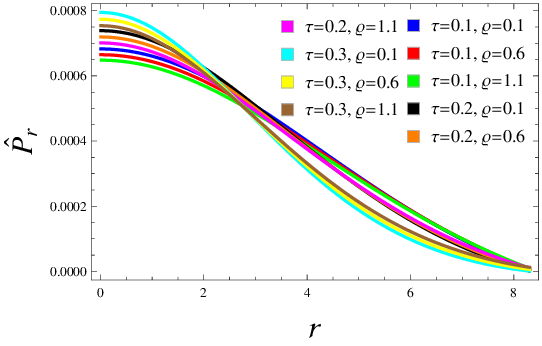,width=0.4\linewidth}
\epsfig{file=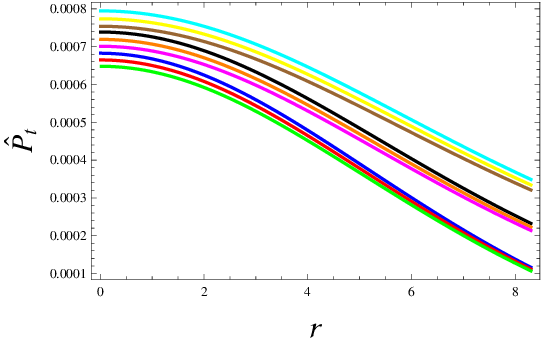,width=0.4\linewidth}\epsfig{file=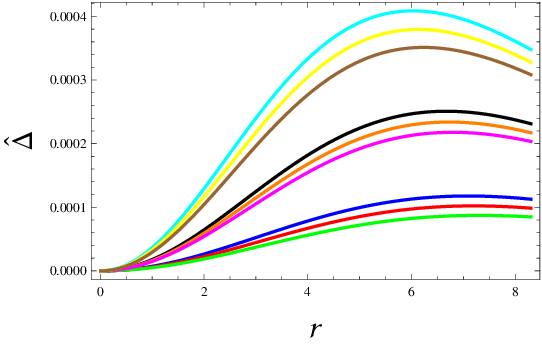,width=0.4\linewidth}
\caption{Fluid parameters for model 1.}
\end{figure}
\begin{figure}[h!]\center
\epsfig{file=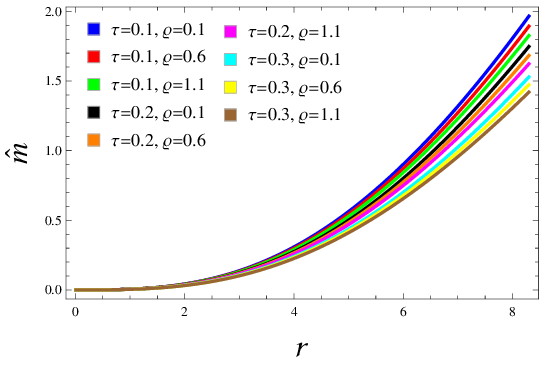,width=0.4\linewidth}\epsfig{file=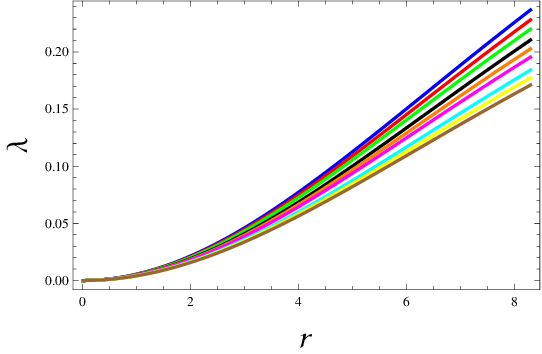,width=0.4\linewidth}
\epsfig{file=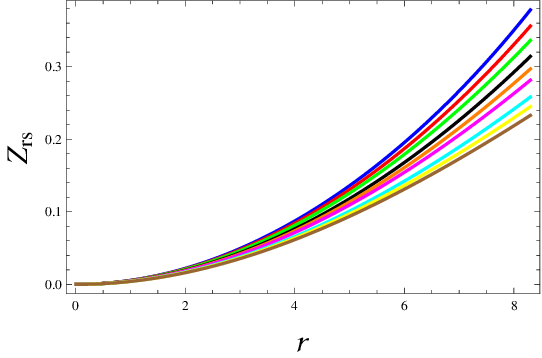,width=0.4\linewidth} \caption{Physical
parameters for model 1.}
\end{figure}
\begin{figure}[h!]\center
\epsfig{file=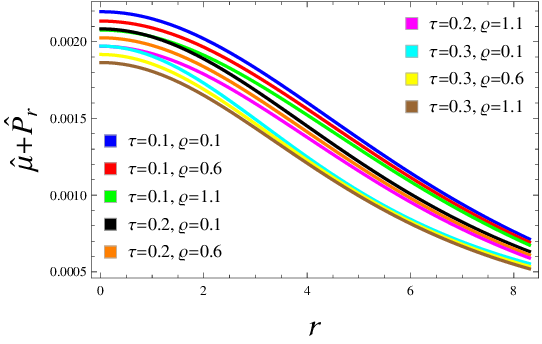,width=0.4\linewidth}\epsfig{file=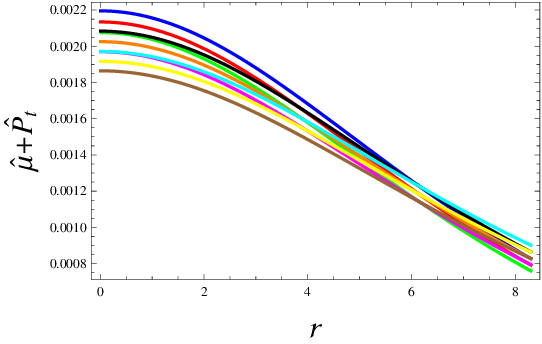,width=0.4\linewidth}
\epsfig{file=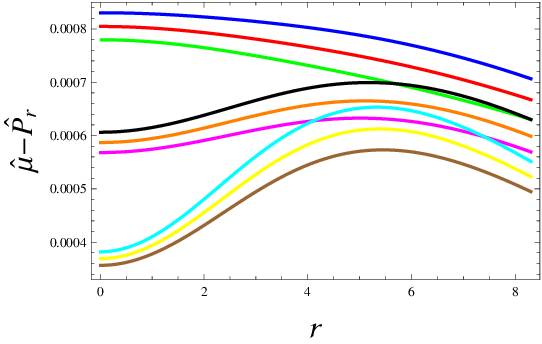,width=0.4\linewidth}\epsfig{file=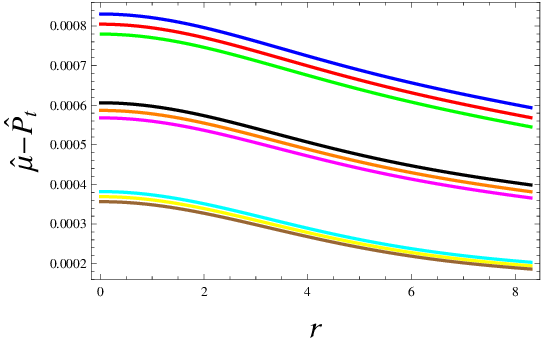,width=0.4\linewidth}
\epsfig{file=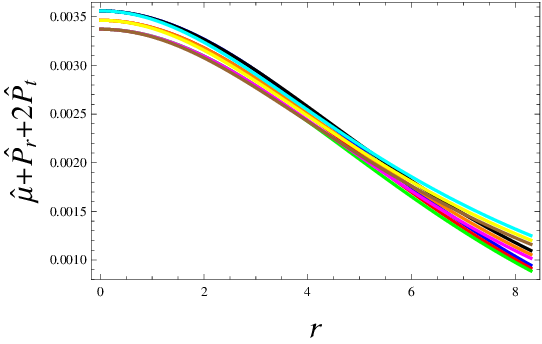,width=0.4\linewidth} \caption{Energy
conditions for model 1.}
\end{figure}
\begin{figure}[h!]\center
\epsfig{file=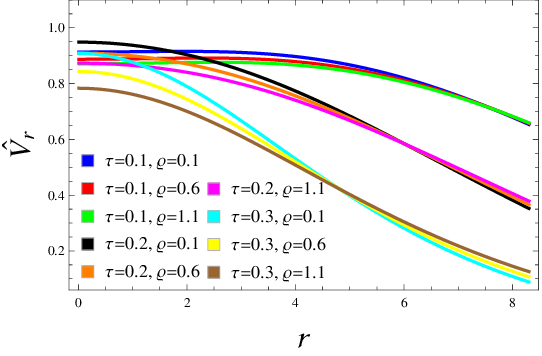,width=0.4\linewidth}\epsfig{file=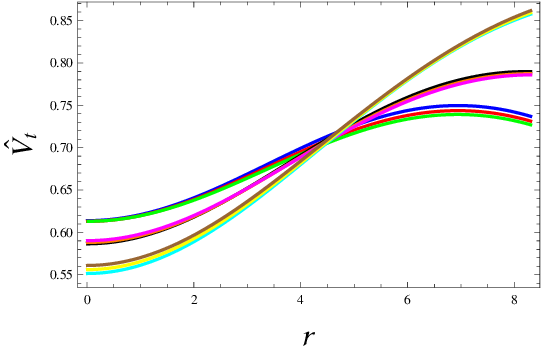,width=0.4\linewidth}
\epsfig{file=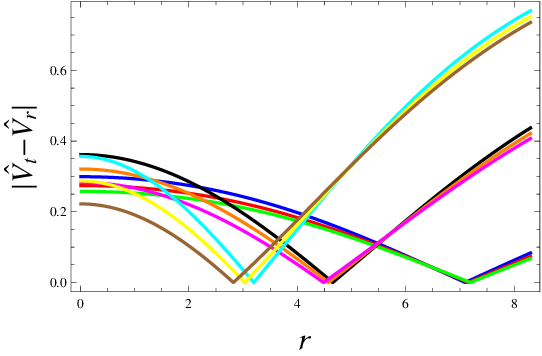,width=0.4\linewidth} \caption{Stability
for model 1.}
\end{figure}
\begin{figure}[h!]\center
\epsfig{file=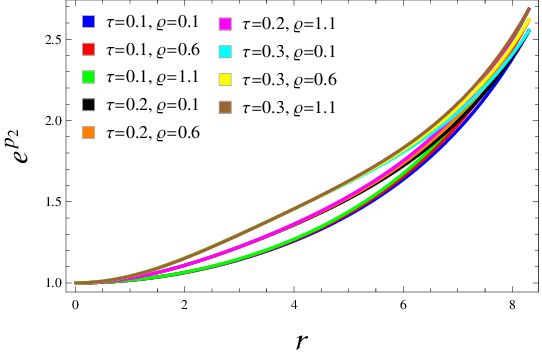,width=0.4\linewidth}\epsfig{file=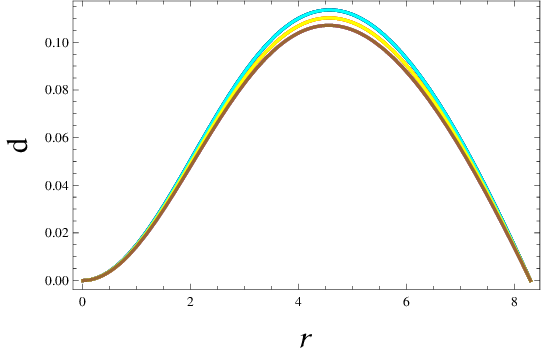,width=0.4\linewidth}
\caption{Modified $g_{rr}$ potential and deformation function for
model 2.}
\end{figure}
\begin{figure}[h!]\center
\epsfig{file=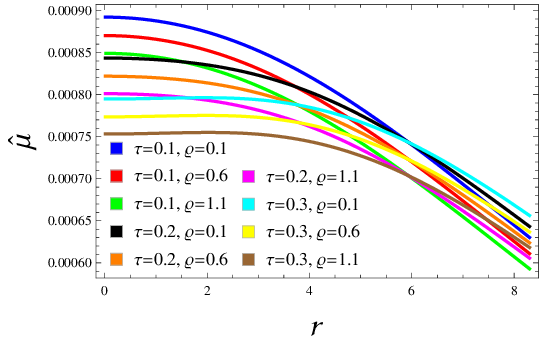,width=0.4\linewidth}\epsfig{file=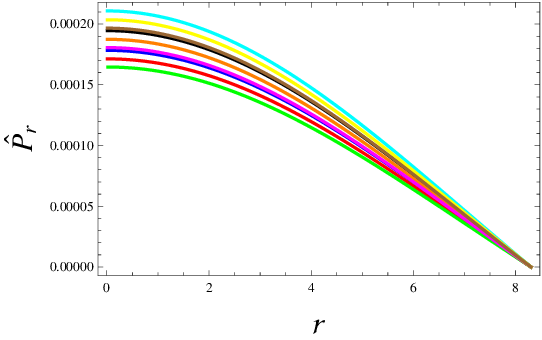,width=0.4\linewidth}
\epsfig{file=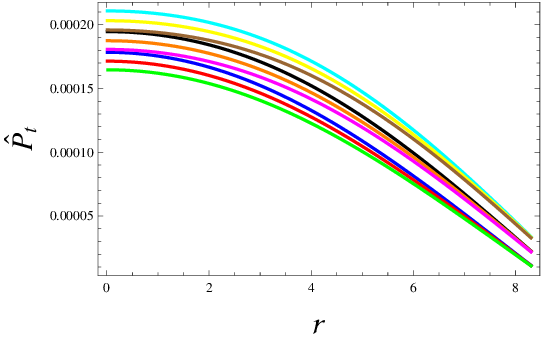,width=0.4\linewidth}\epsfig{file=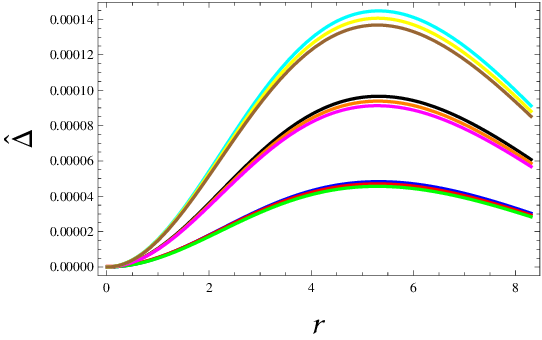,width=0.4\linewidth}
\caption{Fluid parameters for model 2.}
\end{figure}
\begin{figure}[h!]\center
\epsfig{file=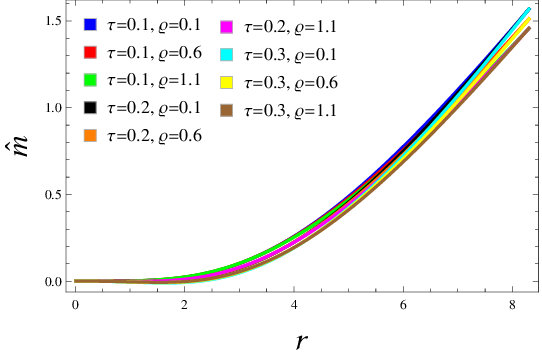,width=0.4\linewidth}\epsfig{file=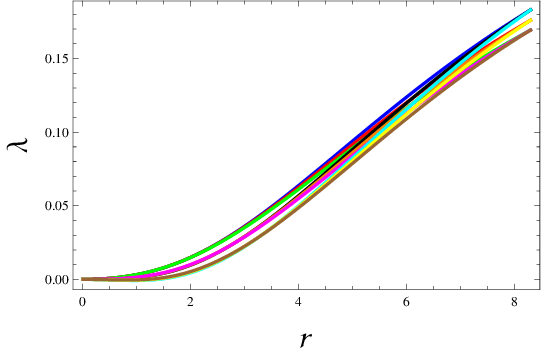,width=0.4\linewidth}
\epsfig{file=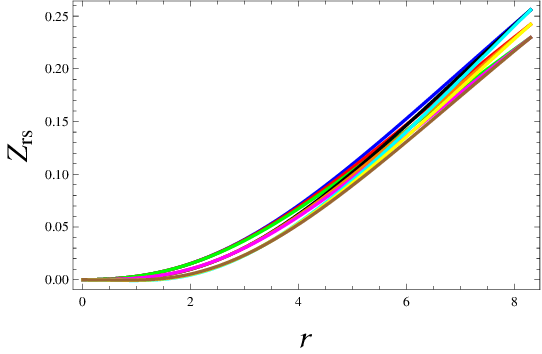,width=0.4\linewidth} \caption{Physical
parameters for model 2.}
\end{figure}
\begin{figure}[h!]\center
\epsfig{file=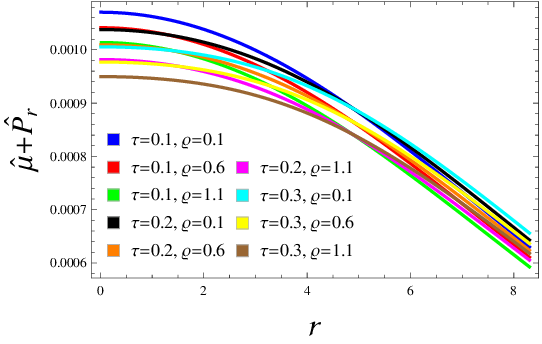,width=0.4\linewidth}\epsfig{file=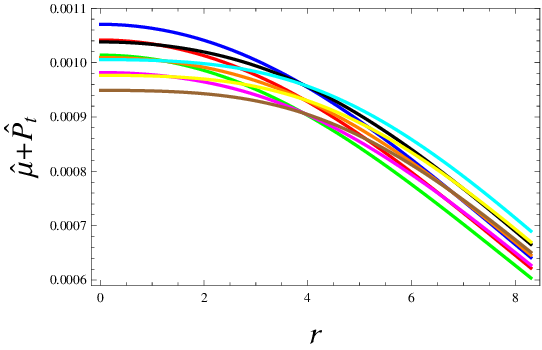,width=0.4\linewidth}
\epsfig{file=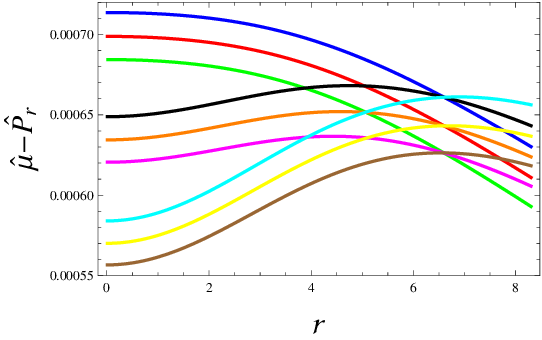,width=0.4\linewidth}\epsfig{file=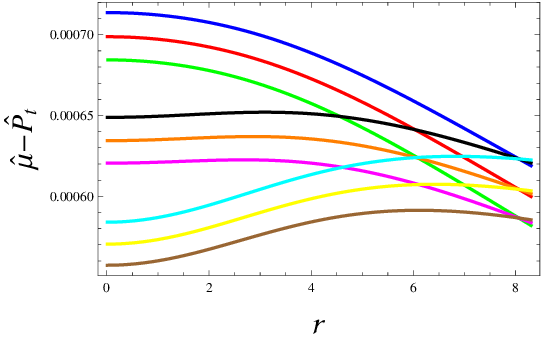,width=0.4\linewidth}
\epsfig{file=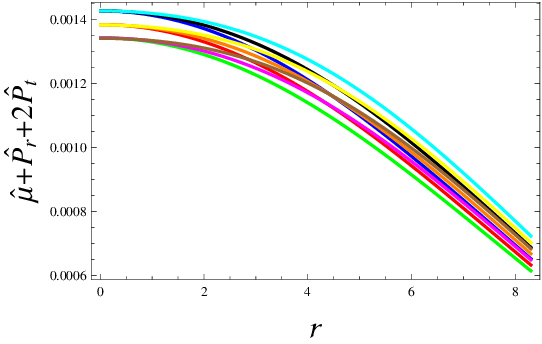,width=0.4\linewidth} \caption{Energy
conditions for model 2.}
\end{figure}
\begin{figure}[h!]\center
\epsfig{file=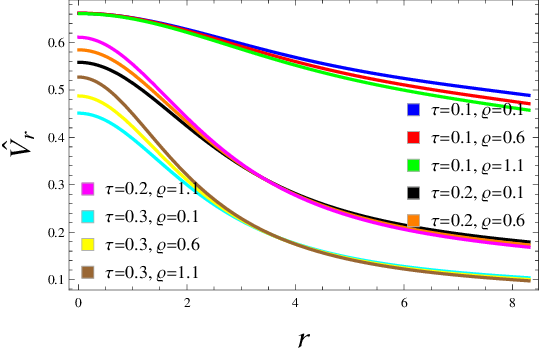,width=0.4\linewidth}\epsfig{file=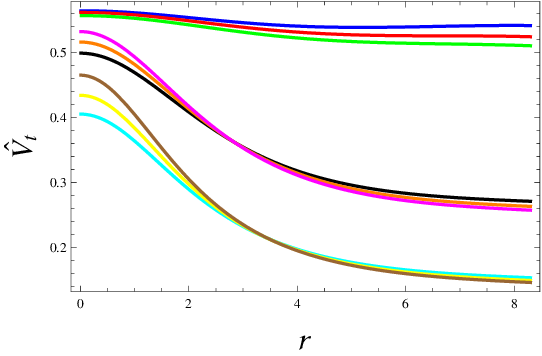,width=0.4\linewidth}
\epsfig{file=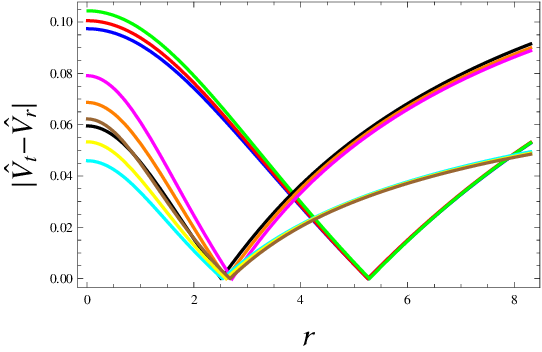,width=0.4\linewidth} \caption{Stability
for model 2.}
\end{figure}

We now put our focus on the behavior of effective physical
parameters such as energy density, pressures in different directions
and anisotropy. We plot all these factors in Figure \textbf{2} which
show that the resulting model 1 can be of physical interest because
it is compatible with the required trend. We observe the impact of
density to be inversely proportional to both parameters $\varrho$
and $\tau$. However, both pressure ingredients show increasing and
decreasing profile for the rise in $\tau$ and $\varrho$,
respectively. Moreover, anisotropy should not be appeared at the
core which is guaranteed from the last plot of Figure \textbf{2}.
Notice that the system with high difference between
radial/tangential pressures is obtained for $\tau=0.3$ and
$\varrho=0.1$. Some other physical factors depending on the mass are
graphically shown in Figure \textbf{3}, from which we find all of
them to be acceptable. Rather than plotting the Misner-Sharp mass in
geometric terms, we plot the mass in combination with the energy
density. This is advantageous because we reach to the conclusion
that the model becomes massive for $\tau=0.1$ and $\varrho=0.1$ as
compared to the other values. The viability analysis is done in
Figure \textbf{4}, exhibiting positive trend of the matter-dependent
factors, hence, model 1 is viable. The three plots of Figure
\textbf{5} show the absence of cracking in the interior, implying
that the solution is stable.

\subsection{Model 2: Pressure-like Constraint}

We now assume that the pressure of both perfect and additional fluid
sources is equal to each other, leading to multiple physically
feasible solutions. We express this as \cite{42babb}
\begin{equation}\label{g55}
P=\Phi_{1}^{1}.
\end{equation}
Combining Eqs.\eqref{g19a}, \eqref{g23} and \eqref{g55} together, a
single equation is obtained in the following
\begin{align}\nonumber
&\frac{\mathrm{d}(r)}{8\pi}\bigg\{\frac{p_1'}{r}+\frac{1}{r^2}\bigg\}
-\frac{e^{-p_2}}{8r^2\big(\varrho^2+6\pi\varrho+8\pi^2\big)}\\\label{g56}
&\times\big[(3 \varrho +8 \pi ) r p_1 '+\varrho  r p_2 '-2 (\varrho
+4 \pi ) \big(e^{p_2 }-1\big)\big]=0,
\end{align}
which becomes when substituting metric components \eqref{g33} and
\eqref{g34} as
\begin{align}\nonumber
&\left(\frac{6 b_2}{b_2 r^2+1}+\frac{1}{r^2}\right)
\mathrm{d}(r)+\frac{6 \pi b_2}{(\varrho +2 \pi ) (\varrho +4 \pi )
\psi ^3 \big(b_2 r^2+1\big)^2} \\\nonumber &\times\big[2 \pi  b_3-3
(\varrho +2 \pi ) \psi+b_2^2 r^4 \big\{8 (\varrho +3 \pi ) \psi +7
(3 \varrho +8 \pi ) b_3\big\}\\\label{g57} &+2 b_2 r^2 \big\{(3
\varrho +11 \pi ) b_3-(5 \varrho +9 \pi ) \psi \big\}\big]=0.
\end{align}
The explicit expression for the deformation function can be easily
obtained for the above equation as
\begin{align}\nonumber
&\mathrm{d}(r)=\frac{-6 \pi  b_2 r^2}{(\varrho +2 \pi ) (\varrho +4
\pi ) \psi ^3 \left(b_2 r^2+1\right) \left(7 b_2 r^2+1\right)}\big[2
\pi b_3+b_2^2 r^4 \big\{8 (\varrho +3 \pi ) \psi \\\label{g58} &+7
(3 \varrho +8 \pi ) b_3\big\}-3 (\varrho +2 \pi ) \psi+2 b_2 r^2
\big\{(3 \varrho +11 \pi ) b_3-(5 \varrho +9 \pi ) \psi \big\}\big].
\end{align}
Equation \eqref{g17} thus produces in conjunction with \eqref{g58}
as
\begin{align}\nonumber
e^{p_2(r)}&=\frac{2 \left(b_2 r^2+1\right)\sqrt{4 b_2
r^2+1}}{\left(2-b_2 r^2\right)\sqrt{4 b_2 r^2+1}-3 b_2 b_3
r^2}-\frac{6 \pi \tau  b_2 r^2\left(7 b_2
r^2+1\right)^{-1}}{(\varrho +2 \pi ) (\varrho +4 \pi ) \psi ^3
\left(b_2 r^2+1\right)}\\\nonumber &\times\big[2 \pi b_3+b_2^2 r^4
\big\{8 (\varrho +3 \pi ) \psi +7 (3 \varrho +8 \pi ) b_3\big\}-3
(\varrho +2 \pi ) \psi+2 b_2 r^2\\\label{g58a} &\times \big\{(3
\varrho +11 \pi ) b_3-(5 \varrho +9 \pi ) \psi\big\}\big].
\end{align}
Equations \eqref{g13} and \eqref{g14} can now be used to construct
the final form of effective fluid parameters. The values of
parameters $\varrho$ and $\tau$ remain the same as we choose earlier
to discuss model 1. However, in this case, $b_2$ is chosen as
$0.002~km^{-2}$. In Figure \textbf{6}, we discuss the behavior of
modified $g_{rr}$ component and deformation function (expressed in
the last two equations), and find them well-behaved. From the right
plot, we see that the function $\mathrm{d}(r)$ reaches a maximum of
$0.11$ at $r=4.5$ and null both at the core and spherical interface
for all choices of parameters. In Figure \textbf{7}, we also check
the trend of fluid parameters and anisotropy in the pressure. We
observe that effective density corresponding to model 2 exhibits
same profile as of model 1 but only near the center of a star. This
relation becomes opposite as we move towards the boundary. The same
observation is carried out regarding the radial component of
pressure. As the anisotropy is concerned, this factor rapidly
decreases near the interface for this resulting model as compared to
the previous one.

In the stellar interior, the mass function and its dependent
properties are also investigated in Figure \textbf{8}. We find the
less massive system in this case when comparing numerical values
with those of model 1 for all values of $\varrho$ and $\tau$. The
existence of usual matter in a celestial body is also needed to
explore because this helps in existing the compact stars physically.
For this, we plot the energy bounds in Figure \textbf{9} and find
all of them lying in positive range, ultimately showing the
viability. Lastly, Figure \textbf{10} exhibits the plots of
stability checkers which are consistent with their criteria, hence,
our model 2 is stable.

\subsection{Advantages of Our Approach over
Existing Methods: A Brief Summary}

To understand how this approach improves upon existing methods for
modeling anisotropic structures in $f(R,T)$ gravity, it is essential
to highlight several key aspects that differentiate this work from
previous studies. Firstly, this methodology employs a gravitational
decoupling approach that allows for a more nuanced interaction
between fluid dynamics and geometry. Traditional models often rely
on isotropic fluid configurations as a starting point, which can
limit the exploration of anisotropic behaviors. By deriving multiple
anisotropic analogs from an established isotropic model, a framework
is introduced that not only retains the foundational characteristics
of isotropic models but also systematically incorporates the
complexities introduced by anisotropic sources. This duality
enhances the robustness of the resulting field equations, enabling a
comprehensive analysis of the matter distribution within compact
objects.

Secondly, the introduction of a new matter source to induce
anisotropic behavior is a significant advancement. Existing models
frequently assume predefined equations of state or specific forms of
matter distribution, which may not accurately reflect the physical
conditions within compact stars. In contrast, this approach allows
for greater flexibility in modeling the internal structure of these
objects by accommodating various forms of anisotropic pressures and
densities. This adaptability is particularly crucial in
astrophysical contexts where the nature of matter can be highly
variable and influenced by factors such as density fluctuations and
phase transitions. Moreover, the use of Heintzmann's ansatz and
specific constraints on additional gravitating sources provides a
structured method to tackle the complexities arising from
anisotropic configurations. This approach facilitates a clearer
delineation between different fluid sources, allowing for a more
straightforward interpretation of the resulting equations. The
ability to split the field equations into two distinct sets not only
simplifies the analysis but also enhances the clarity with which we
can understand the physical implications of anisotropy in stellar
structures. Finally, a thorough assessment of the physical validity
of the models has been conducted using observational data from a
specific star candidate, LMC X-4. This empirical grounding is
crucial as it ensures that the theoretical advancements made through
this approach are not merely abstract but are aligned with real
astrophysical phenomena. By demonstrating that the developed models
adhere to established acceptability criteria under certain
parametric values, a compelling evidence has been provided that this
method offers practical improvements over existing modeling
techniques.

In addressing the limitations of stability tests, it is essential to
consider both causality conditions and Herrera's cracking criteria.
Causality condition ensure that the propagation of signals does not
exceed the speed of light, which is a fundamental aspect of
relativistic theories. Furthermore, Herrera's cracking criteria
provide a framework for analyzing the stability of anisotropic
configurations, particularly in relation to the onset of
instabilities such as cracking or fragmentation within the matter
distribution. We acknowledge that not all configurations may satisfy
Herrera's criteria across all parameter spaces. This limitation
suggests that further investigation is needed to explore the
stability of our models under varying conditions and to identify
potential regions where instabilities could arise. In addition to
these considerations, we also recognize the importance of other
stability criteria that may be relevant in our context. For
instance, examining the effects of perturbations on our solutions
could reveal additional insights into their stability. However, this
is not performed in our analysis at this time.

Inflation potentials play a crucial role in the context of
cosmological models, particularly in explaining the dynamics of the
early universe. In essence, inflation refers to a rapid exponential
expansion that occurred shortly after the Big Bang, driven by a
scalar field known as the inflation. The potential energy associated
with this field dictates the rate of expansion and the dynamics of
the universe during this phase. In relation to our model derived
from isotropic fluid dynamics, inflation potentials are significant
because they provide a framework for understanding how anisotropic
behavior can emerge from an initially isotropic configuration. By
introducing a new matter source that influences the gravitational
dynamics, we can explore how these potentials affect the stability
and evolution of anisotropic models. Moreover, the relationship
between inflation potentials and our model is evident in how these
potentials can influence the effective equations governing
fluid-geometry interactions. The anisotropic configurations we
develop can be viewed as modifications to standard cosmological
models, incorporating effects from inflationary dynamics.

\section{Concluding Remarks}

In this study, we have developed various anisotropic analogs of a
well-known isotropic ansatz within the framework of the model
$f(R,T)=R+2\varrho T$. Initially, we assumed a spherical
configuration filled with isotropic perfect fluid, which transitions
to the anisotropic state upon introducing an additional source at
the action level. The modified action \eqref{g1} consequently leads
to field equations that incorporate the influences of both the
original isotropic configuration and the additional sources within
the modified theory. To manage these equations effectively, we
employed the MGD strategy to split them into two distinct sets. The
first set of equations was addressed by adopting the Heintzmann's
solution as follows
\begin{align}\nonumber
e^{p_1(r)}&=b_1^2 \left(b_2 r^2+1\right)^3,\\\nonumber
e^{p_2(r)}&=\frac{2 \left(b_2 r^2+1\right)\sqrt{4 b_2
r^2+1}}{\left(2-b_2 r^2\right)\sqrt{4 b_2 r^2+1}-3 b_2 b_3 r^2},
\end{align}
and derived the constant triplet $(b_1,b_2,b_3)$ by applying smooth
matching criteria at the junction $\Sigma:~r=\bar{R}$. Additionally,
the $\Phi$-sector described by Eqs.\eqref{g21}-\eqref{g23} involved
four unknowns that required determination. Subsequently, we imposed
distinct constraints on $\Phi_{\sigma\omega}$, leading to the
development of two distinct solutions. These solutions were then
combined via the parameter $\tau$ to incorporate contributions from
both sources, resulting in the generation of novel anisotropic
analogs.

We have further explored the criteria for acceptability which ensure
that the derived solution must align with these conditions if it is
claimed to be physically relevant. The physical attributes for all
constructed models have been visually interpreted for specific
parameter values, such as $\varrho=0.1,0.6,1.1$ and
$\tau=0.1,0.2,0.3$. Our observations reveal consistent behavior of
fluid triplet across all parametric combinations. Also, the
anisotropic factor has been noticed to be an increasing function of
$r$. The mass functions were plotted for every developed model to
see which condition on the additional fluid source produce more
dense interior. We observe model 1 to be more dense in comparison
with the other model. Moreover, some other factors have also been
demonstrated to be consistent with the observational data. The
energy bounds were seen to be positive, indicated the viability of
all models. Finally, occurrence of cracking has not been observed in
any case, hence, models 1, 2 and 3 are stable. Furthermore, when
compared to prior studies within $f(R,T)$ gravity, our models offer
unique insights into the physical validity of anisotropic
configurations. Previous works have explored various aspects of
anisotropic stars and their stability under different gravitational
frameworks; however, our approach specifically emphasizes the
transformation of isotropic solutions into anisotropic ones while
maintaining consistency with the underlying physics. We must
highlight here that the anisotropic Heintzmann's extensions are more
suitable as compared to those of Buchdahl's solution \cite{25ac}. We
can find all these results in GR when putting $\varrho=0$.\\\\
\textbf{Data Availability Statement:} This manuscript has no
associated data.

\end{document}